\documentstyle[epsf,eqsecnum,prb,aps]{revtex}
\begin{document}
%\draft  \preprint{}

%\twocolumn[\hsize\textwidth\columnwidth\hsize\csname @twocolumnfalse\endcsname
\title{\bf Influence of Point Defects on the Shear Elastic Coefficients and on 
the Melting Temperature of Copper}

\author{Amit Kanigel, Joan Adler and Emil Polturak} 

\address{Department of Physics, Technion-IIT, Haifa 32000, Israel.} \date{\today} \maketitle 

\begin{abstract}
We present molecular dynamics simulations of the influence of point defects on 
the shear elastic coefficients of copper. We find that vacancies do not influence
these coefficients at all, while the introduction of interstitials causes a 
large reduction in the elastic coefficients. The simulations establish a phase 
diagram of the melting temperature as a function of the density of interstitials.
 A crystal having no free surface undergoes bulk mechanical melting as a result
 of the vanishing of 
$C^{'} \equiv (C_{11}-C_{12})/2$ once the specific volume reaches a critical value,
 equal to the experimental volume of liquid phase.  
This critical volume is history independent, in the sense that it can be
reached either by heating the crystal or by adding defects at a constant 
temperature.
These results generalize the Born model of melting for the case where point
 defects are present.

\end{abstract}


]
\section{INTRODUCTION}
Single phase models of bulk melting  
usually treat this process as a destabilization
of the solid due to proliferation of crystalline defects or excitations. 
Among these, Lindemann suggested that melting occurs when the amplitude 
of atomic vibrations exceeds a critical value\cite{ubb}. Born proposed that
melting is connected with
the vanishing of the shear elastic coefficients \cite{born}. Other models
suggest that melting is a consequence of the 
proliferation of intrinsic crystalline defects such as 
vacancies \cite{frenkel}. However, there is no clear experimental
evidence supporting any one of these models, all of which rely on a single
 type of defect or excitation. A model of melting involving
interstitials coupled to shear strains proposed by Granato \cite{granato} is
 an exception as it involves coupling of defects and excitations.

Recently, a dramatic decrease of the shear resistance was observed near 
the bcc-hcp transition
of solid $\rm ^4He$\cite{berent}. These observations could have been
 interpreted as 
resulting from a coupling
of point defects with a phonon which softens near the 
transition. Were it not for the bcc-hcp transition, one could envisage
a continuation of this process until the crystal would melt. 
Thus, we have experimental evidence of a microscopic process which 
can potentially lead to melting. However, this scenario requires 
a coupling of phonons and point defects, and therefore supports
a picture similar to Granato's, in which more than one type of excitation 
is involved.
To understand this further, it is necessary to investigate how
point defects couple to phonons, and how this influences the 
shear resistance. In the following,  
we report results of molecular 
dynamics (MD) simulations of the influence of point defects on the elastic properties
of copper. Copper was chosen because of the large body of experimental data to 
which the results of the simulations can be compared. We focus on the shear
elastic coefficients, with the intention to test the Born model in the 
presence of point defects.   

\section{SIMULATION METHOD}

We performed molecular dynamics simulations on samples containing up to
2048 atoms using  
the ``Tight-Binding (TB) Potential'' with parameters calculated by Cleri and 
Rosato \cite{cleri} appropriate to copper. Newton's equations were solved 
using Gear's Predictor-Corrector algorithm \cite{allen}.
The time-step used was about 1 fSec.
For temperature control we used the method known as the Nose-Hoover thermostat \cite{hoover1}.
The method of Berendsen \cite{berendsen} was used to calculate the volume 
of the samples. This method does not generate trajectories in the constant 
NPT (here, N denotes the constant number of atoms, 
P is the constant pressure, and T is the temperature) ensemble, but it sets the mean value of the pressure to the desired 
one while predicting the correct volume of the system.
For ensemble average calculations we used the Parrinello and Rahman (PR) method \cite{par_ram1}.  
  The statistical ensemble created by the Parrinello Rahman
combined with Nose's thermostat was identified as the
isothermal-isotension (NtT) ensemble of statistical mechanics
\cite{ray&rahman1,ray&rahman2}(here,  
t is the constant tension).
We calculated the elastic constants using a fluctuation formula in
 both the canonical and in the NtT ensembles. 
  
In the NtT ensemble the formula for the isothermal elastic constants $C_{ijkl}$ is 
\begin{equation}
\delta(\epsilon_{ij} \epsilon_{kl}) = k_{B}TC_{ijkl}^{-1}/V_{0}
\end{equation}
where $\epsilon$ is the strain 
and $V_{0}$ is the volume of the sample at zero strain.
Calculating elastic constants using this formula involves two steps.
First we have to find the reference matrix, $h_{0}$, defining the shape
 of the simulation box. The condition which determines the shape of the
box is that the sample is at zero stress. 
After running the simulation to calculate $h_{0}$, the program should be used
again to calculate the fluctuations.
Not many studies have been made using this method because of the very slow 
convergence of the strain fluctuations \cite{ray,palladium}.

 In the constant NVT ensemble (here, N denotes the constant number of atoms, 
V is the constant volume, and T is the temperature), MD can determine the elastic constants by using 
fluctuation formulas derived by Ray and Rahman \cite{ray&rahman1}:
\begin{equation}
C_{ijkl}=-\frac{V_{0}}{k_{B}T} \delta( \sigma_{ij} \sigma_{kl}) + 
\frac{2 N K_{B} T}{V_{0}}(\delta_{ik} \delta_{jl}+\delta_{im} \delta_{jk}) + 
<B>
\label{NVT}
\end{equation}
where $\delta_{ij}$ is the Kronecker delta and $<B>$ is the ensemble
average of the Born term, defined as
\begin{equation}
B=(\partial^{2}V/\partial r_{ab}^{2} - (\partial V/ \partial r_{ab})/r_{ab})r_{abi}r_{abj}r_{abk}r_{abl}
\label{Born_term}
\end{equation}
where $ijkl$ are the indices of the elastic constants tensor.
This formula for calculating the elastic constants is much less
compact than the NtT formula, mainly because it contains the explicit
second derivative of the inter atomic potential. The main advantage of this
method, and the reason for it being used more than the NtT method, is
the faster convergence of the stress-stress fluctuation term relative to
the convergence of the strain-strain fluctuation term.
Two steps are required to calculate the elastic constants using this formula.
 As in the NtT formalism first one has to find the zero
stress reference matrix and volume $h_{0}$ and $V_{0}$ and then run
the NVT MD program to calculate the elastic constants using equation
(\ref{NVT}).

\section{Validation and comparison with experiment}
In order to validate the programs and to check the potential at various 
conditions we performed a series of calculations of the physical properties of
copper using our programs and compared the results with experimental data.

\begin{figure}
\centerline{\epsfysize=7.0cm \epsfbox{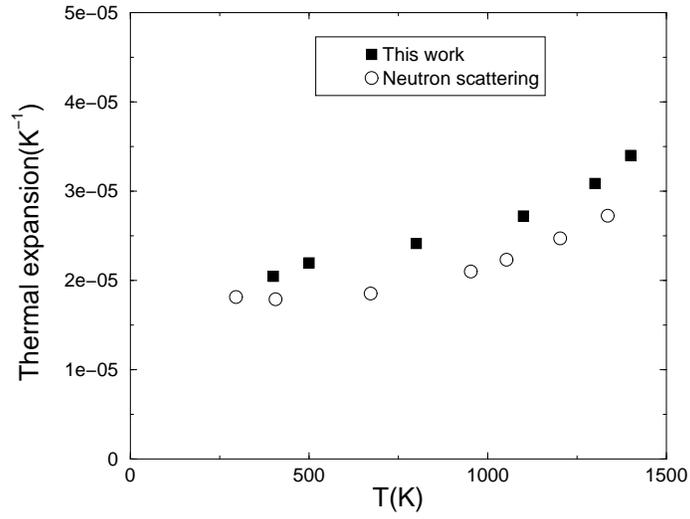}}
\caption{Thermal expansion coefficient of copper as function of temperature 
obtained from our simulations compared with data determined by a neutron 
scattering experiment
 {\protect \cite{larose}}.}
\label{alpha}
\end{figure}

First, we calculated the thermal expansion up to the melting point at zero 
external pressure. Our results are compared with neutron scattering 
data \cite{larose} in Fig. \ref{alpha}. 
As expected, the calculated values are close to the experimental values
at low temperatures (We must bear in mind that the lattice
constant at zero temperature was used as input in the fitting of the
potential). The agreement remains good even at elevated temperatures, with
a maximal difference of about 10\%.

Secondly, the phonon spectrum,  $\omega (\vec{k})$ , was determined by 
calculating the time Fourier-transform of the following 
quantities\cite{caprion}:
\begin{equation}
f_{\alpha}(\vec{k},t)=\sum_{i=1}^{N} (\vec{u}_{\alpha} \cdot \vec{v}_{i}) \cos(\vec{k} \cdot \vec{r}_{i})
\end{equation}
where, $\vec{r}_{i}$ and $\vec{v}_{i}$ are the time dependent positions
and velocities of the $i$th atom. The polarization vectors are defined
as $\vec{u}_{l}=\vec{k}/k$ in the longitudinal case, while
$\vec{u}_{t1}$ and $\vec{u}_{t2}$, the transverse polarizations, are
perpendicular to $\vec{u}_{l}$ and to each other.

\begin{figure}
\centerline{\epsfysize=7.0cm \epsfbox{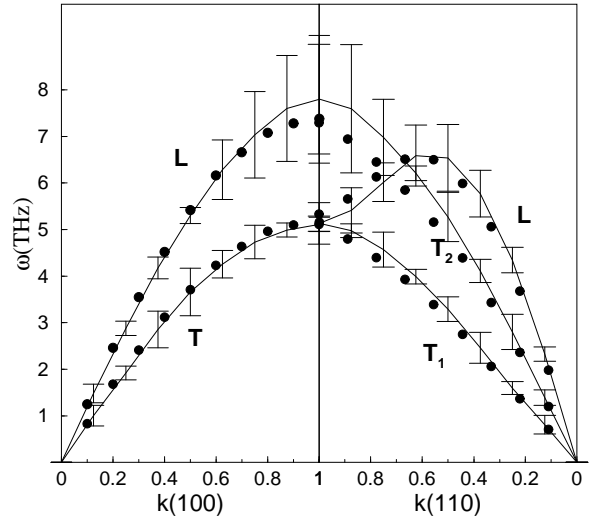}}
\caption{Phonon dispersion curve at 400K along (100) and (110).
Solid lines are the result of the simulations, while
 the circles are 
experimental results at 80K {\protect \cite{nilsson}}.}
\label{WvsK}
\end{figure}

We then calculated the phonon spectrum along the $(100)$ and $(110)$
directions.  Due to the importance of long-time correlations the
calculations were made under conditions of constant energy without
temperature control. We ran the MD program for $20000$ time steps with
temperature and pressure control, $10000$ time steps with temperature
control only, and then $f_{\alpha}(\vec{k},t)$ was saved to disk during
$200000$ time steps under energy conservation conditions. The resulting 
spectrum, shown in Fig. \ref{WvsK}, is in good agreement with neutron
scattering data\cite{nilsson}.

\begin{figure}
\centerline{\epsfysize=6.0cm \epsfbox{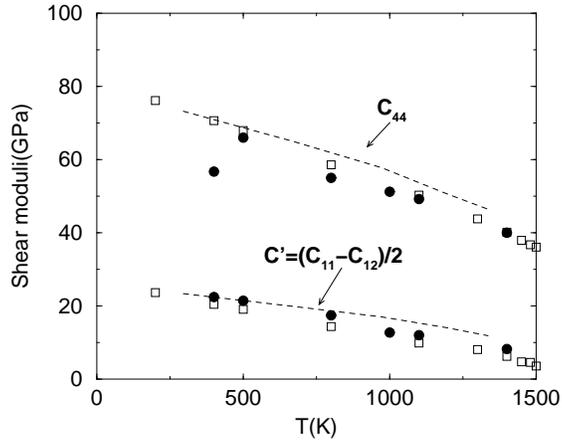}}
\caption{Variation of the shear moduli with temperature: PR method, 
$\bullet$; NVT MD, $\Box$; experimental data {\protect \cite{larose}}, dashed line.}
\label{shear_vs_temp}
\end{figure}

Thirdly, to test the algorithms used to obtain the elastic constants,
we calculated their variation with temperature 
using both approaches described in section II.
The system was initially equilibrated for $50000$ steps, and $300000$
subsequent
steps were used to calculate the fluctuations. We found that the convergence 
of the PR method was slower than that of the NVT method. 
To monitor the
convergence of the algorithm, we compared values of the elastic coefficients
calculated along directions which are symmetrically equivalent.
In the final result, after $300000$ time steps, in both methods, the three 
equivalent elastic constants differ by no more than $10\%$. From this variance,
one can estimate the accuracy of the calculation.
Comparison with the experimental values measured by neutron
scattering \cite{larose}, shows that the shear moduli of the model-lattice 
and of real copper soften in the same way within the estimated accuracy of the
simulations (e.g. 10\%). The results are shown in Fig. \ref{shear_vs_temp}. 

The typical computer time needed to calculate the elastic constants varies 
between 4 days for a calculation in the NVT ensemble to one week in the NtT 
ensemble on a Pentium-Pro computer. 

\section{Influence of point defects}
After successful validation of the pure system,
point defects were introduced into the lattice by either removing an
atom (vacancy) or adding one (interstitial). The simulations were done using a
box having a fixed number of sites. Periodic boundary conditions 
were used throughout.
In order to prevent mutual annihilation of the defects, we used defects of 
only one species in each run.  
We first used simulated annealing\cite{adler1} to find the most
stable configuration of a vacancy and of an interstitial in the
model.  The formation energy of a vacancy was found to be $1.27{\rm
eV}$. The lowest energy configuration of an 
interstitial atom was found to be a (100)-split interstitial
with a formation energy of $3.28{\rm eV}$. These formation energies agree well
with experimental data \cite{wollenberger}, and with previous
simulations \cite{cleri}. Formation energies of other possible 
configurations of interstitials were determined using the conjugate
gradient method, and are summarized in Table \ref{ein_shem}. We point out that
the difference between these energies is on the order of 0.1 eV ($\sim
$ 1000K), 
which means that interstitials of various configurations may be present at
high temperatures.

\begin{figure}
\centerline{\epsfysize=6.0cm \epsfbox{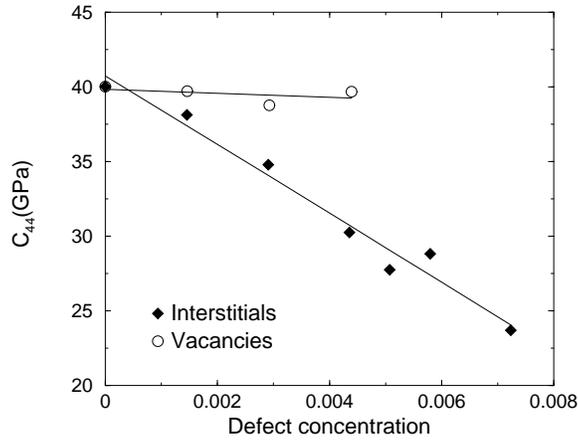}}
\caption{Dependence of $C_{44}$ on the concentration of defects at $T=1400K$,
 showing the difference between the influence of vacancies and interstitials.
 }
\label{C_vs_vac_and_int_1400}
\end{figure}

The influence of point 
defects was determined by calculating the shear elastic coefficients
in samples containing different concentration of defects using MD.
For any concentration of defects, the number of atoms and unit cells in the 
sample is kept constant, so that the number of externally introduced defects 
is conserved. Under this constraint, atoms and defects are free to diffuse, 
agglomerate etc.
We first discuss vacancies.  We found that the influence of 
vacancies on the shear elastic coefficients is too small to be resolved 
within our stated accuracy.
For example Fig. \ref{C_vs_vac_and_int_1400} shows that $C_{44}$ is independent
of the concentration of vacancies.  
This is correct both at low temperatures and near the 
melting point.  Previously, influence of vacancies on the 
elastic properties at zero temperature was treated analytically by 
Dederichs, et al. \cite{dederichs2}
who also reviewed earlier work. The two body potentials used by these
authors are less accurate than the one used here. They found a reduction
of 2-3 \% in the values of $C_{44}$ and 1.5-3 \% in values of $C^{'}$ per 
percent of vacancies. The effect is indeed small, falling within the uncertainty
limits of our calculation. 

We turn now to the influence of interstitials. In contrast to the vacancies,
interstitials seem to cause a large softening of of the shear moduli,
(see Fig. \ref{C_vs_vac_and_int_1400}). At low temperatures,
we can qualitatively compare the results with the analytical 
calculation of 
Dederichs {\it et al}, made at T=0K \cite{dederichs}, and with the
irradiation experiment of Holder {\it et al}\cite{holder},
made at T=4.2K. In the latter, the defects produced by the irradiation are 
Frenkel pairs and
not interstitials. The common trend is that 
defects induce a large softening of the shear moduli. The softening is
anisotropic, meaning that  
$C_{44}$ softens more than $C^{'}$. Bearing in mind the different 
conditions under which the different studies were made, the qualitative 
agreement can be regarded as good.

The main new thrust of this work regards studying the influence of 
point defects on  
the shear moduli at increasingly higher temperatures, up to the melting 
temperature $T_M$. We emphasize that due to the use of periodic boundary 
conditions, the
crystal can be super-heated and $T_M$ is higher than the thermodynamic 
melting temperature.

\begin{figure}
\centerline{\epsfysize=6.0cm \epsfbox{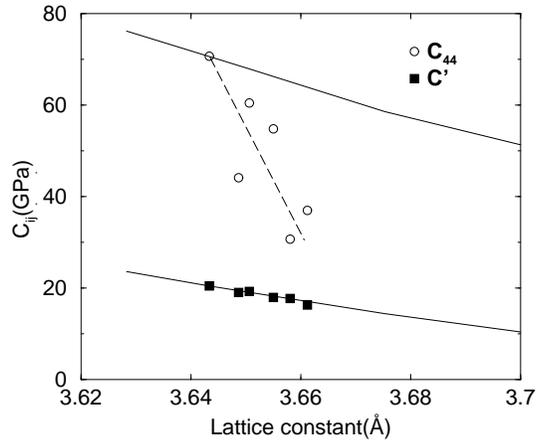}}
\caption{The variation of the shear moduli vs. the lattice constant. The solid 
line represents thermal dilation and the symbols represent dilation due to 
interstitials. The dashed line is a guide to the eye.}
\label{CvsV400}
\end{figure}

In general, our results can be explained by the combined effect of volume change 
and the dia-elastic effect\cite{dederichs3}. Volume changes resulting from 
introduction of defects arise
due to the specific volume per defect being larger than that of an atom
residing at a normal lattice site. We can compare the change of the elastic
coefficients caused by defects with that induced by heating the crystal up
to a temperature where thermal expansion increases the volume by the same 
amount. If the dependence of the elastic coefficients on the volume is the same,
it would mean that effect of the defects is simply to expand the lattice.
We demonstrate such a comparison in Fig. \ref{CvsV400} for data obtained
at T= 400K.
As one can see in this figure, 
the change of $C^{'}$ due to thermal expansion (without defects) is identical
with that induced by increasing density of defects at a constant temperature 
when plotted against the lattice constant. 
This proves that $C^{'}$ is affected only by volume changes. On the other
hand, the change of $C_{44}$ due to thermal expansion is very different from
that induced by increasing the density of defects, and cannot be explained 
by the volume increase alone. 
We attribute this behavior to the dia-elastic effect.  
Briefly, the dia-elastic effect describes the coupling of 
the internal vibration modes of the interstitial with the phonons.
Some of the internal vibration modes,
the so called resonance modes, have frequencies within the band of the usual
lattice spectrum. Mode coupling will lower the long wavelength part of the 
phonon branch related to the elastic coefficients. For example,
the (100) split interstitial is coupled to (100) shear strain. If the 
lattice is deformed in this way the
response of the atoms forming the defect would be larger than the response of
atoms occupying normal lattice sites. The large response leads to a reduction 
of $C_{44}$ elastic coefficient\cite{dederichs,dederichs2}. 
On the other hand, $C^{'}$ is associated with (110) shear strain, which 
does not couple to the (100) split interstitial. Consequently, 
the decrease of $C^{'}$ is induced only by volume expansion, as indeed seen in
Fig.\ref{CvsV400}.

Our results indicate that at high temperatures we can make one of 
the shear moduli vanish by adding a sufficient number of
self-interstitials to the sample.  
The number of defects needed for this purpose is temperature 
dependent. The absolute change of the shear moduli per \% of interstitials is
larger for $C_{44}$ at all temperatures, as summarized in 
Table \ref{sum1400}. 

\begin{figure}
\centerline{\epsfysize=6.0cm \epsfbox{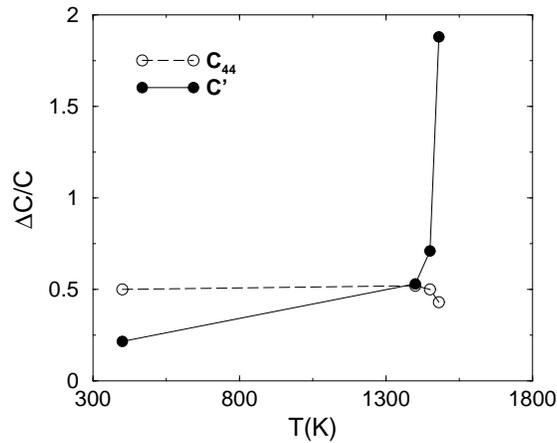}}
\caption{Relative softening of the shear moduli per percent of interstitials 
as function of temperature. The lines are guides to the eye.}
\label{delta_C_rel}
\end{figure}

However, the {\em relative} sensitivity of
$C^{'}$ becomes larger at high temperatures as melting is approached.    
because its
initial value at elevated temperatures is already very small.  
This is  plotted in Fig. \ref{delta_C_rel}. 
Consequently,$C^{'}$ is the first one to vanish. This result is quite
surprising, as it is opposite to what one would expect based on the
low temperature behaviour\cite{holder}. 

\begin{figure}
\centerline{\epsfysize=6.0cm \epsfbox{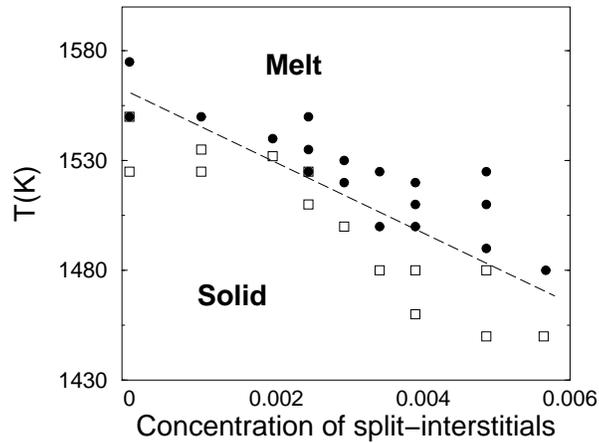 }}
\caption{The influence of self interstitials on the melting temperature
obtained using periodic boundary conditions (a crystal without free surfaces).}
\label{phase_dia_int}
\end{figure}

We determined the melting temperature within our model
at different temperatures and defect concentration, by monitoring
the structural order parameter after $200000$ time steps.  
The results are summarized in Fig. \ref{phase_dia_int}.
Each point in the figure represents the results of several runs
made  at constant T and with the same defect density, but at a
different random initial configuration.  
The melting process found is mechanical melting, and the sample
collapses due to the vanishing of $C^{'}$.
This was found by correlating the temperatures at which $C^{'}$ vanishes
with the melting temperature. The melting process is
rather fast, its duration being on the order of a typical vibration
time of the crystal. Thus, this is a bulk melting mechanism.  As can
be seen from our results the interstitials reduce the mechanical
melting point,and a concentration of about $0.005$ lowered the melting
point by about $\rm 60^{o}K$.  The same cannot be said about vacancies,
which  did not reduce the mechanical
melting point even at very high concentration (up to 4\%).

 It turns
out that the specific volume per atom at the phase boundary
in Fig. \ref{phase_dia_int} is fixed 
at 13.31$\rm \AA^3$. The same specific volume is found by extrapolating the 
dependence of $C^{'}$ on the volume to the value at which $C^{'}$=0 . 
This specific value is consistent with the specific volume of the liquid 
phase at T=1356K, the thermodynamic melting temperature,
as suggested previously\cite{wolf,tallon}.
The way
one reaches that specific volume, either by heating or by adding 
defects at a constant temperature does not seem to matter. 

In summary, we have simulated the influence of point defects on the 
shear moduli of Cu at temperatures up to $T_m$. 
Mechanical melting occurs due to the fullfillment of the Born criterion,
where one of the shear moduli of the solid vanishes. We found
that mechanical melting of the crystal is brought about by the vanishing
of $C^{'}$, rather than that of $C_{44}$. 
Melting occurs once the specific
volume reaches a critical value. This value is independent of the path by
which it is reached, either by thermal expansion alone, or by any combination
of thermal expansion and of expansion due to interstitials introduced into 
the sample. Since the specific volume per vacancy is smaller than the atomic
volume, vacancies do not expand the crystal. Hence, vacancies play no role
in bulk melting.

\section{Acknowledgements}
We are grateful to David Saada, Adham Hashibon, Zaher Salman for their
contribution to this project, and IUCC for the use of the computer facilities.
We thank R. W. Cahn for his comments.
This work was supported in part by the
Israel Science Foundation and by the VPR Technion Fund for the Promotion
of Research.

\begin{table}
\begin{center}
\begin{tabular}{|c|c|} \hline
Configuration & $E^f(eV)$ \\ \hline
vacancy & 1.27 \\
(100) split & 3.28 \\
Octahedral & 3.45 \\
(110) split & 3.55 \\
(111) split & 3.78 \\ \hline
\end{tabular}
\caption{Formation energies of various point defects.} 
\label{ein_shem}
\end{center}
\end{table} 

\begin{table}
%\begin{center}
\begin{tabular}{|c|c|c|} \hline
Temp(K) &  $\Delta C_{44}(GPa / \% inters.)$ & $\Delta C^{'} (GPa / \% inters.)$ \\ \hline
400   & $-39 \pm 13$  & $-4.5 \pm 2.5$  \\ 
1400  & $-23 \pm 1.7$ & $-3.5 \pm 2.0$  \\
1450  & $-18 \pm 2.5$ & $-4.1 \pm 2.0$  \\
1480  & $-15.5 \pm 1.3$ & $-7.8 \pm 3.5$  \\ \hline
\end{tabular}
\caption{Change in shear moduli per percent of self-interstitials at various temperatures.} 
%\end{center}
\label{sum1400}
\end{table} 

\vspace{1.0cm}

\end{document}